\documentstyle{mn}

\newif\ifAMStwofonts
\def\lesssim{\mathrel{\hbox{\rlap{\hbox{\lower4pt\hbox{$\sim$}}}\hbox{$<$}}}}

\def\gtrsim{\mathrel{\hbox{\rlap{\hbox{\lower4pt\hbox{$\sim$}}}\hbox{$>$}}}}

\def\msun{$M_{\odot}$~}

\def\ll_lsun{$\log{L/\rm L_{\odot}}$~}

\def\masa_msun{$M/ \rm M_{\odot}$~}

\def\m_mstar{$M/M_{*}$~}

\def\mean#1{{\langle}#1{\rangle}}

\title{ Evolution of iron core white dwarfs}

\author[J. A.  Panei, L.  G. Althaus  \& O.  G. Benvenuto] {J. A.
Panei\thanks{Fellow  of  the  Universidad  Nacional  de La Plata,
Argentina.   Email:   panei@fcaglp.fcaglp.unlp.edu.ar},   L.   G.
Althaus\thanks{Member  of   the   Carrera   del   Investigador
Cient\'{\i}fico y Tecnol\'ogico, Consejo Nacional de
Investigaciones
Cient\'{\i}ficas  y   T\'ecnicas  (CONICET),   Argentina.  Email:
althaus@fcaglp.fcaglp.unlp.edu.ar}        and        O.        G.
Benvenuto\thanks{Member   of   the   Carrera   del   Investigador
Cient\'{\i}fico, Comisi\'on  de Investigaciones  Cient\'{\i}ficas
de   la   Provincia   de   Buenos   Aires,   Argentina.    Email:
obenvenuto@fcaglp.fcaglp.unlp.edu.ar}  \\  Facultad  de  Ciencias
Astron\'omicas y Geof\'{\i}sicas, Paseo del Bosque S/N, (1900) La
Plata, Argentina}

\date{June 28}

\pagerange{\pageref{firstpage}--\pageref{lastpage}}

\pubyear{1999}

\begin{document}

\maketitle

\label{firstpage}

\begin{abstract} Recent measurements  by Hipparcos (Provencal  et
al. 1998) present observational evidence supporting the existence
of  some  white  dwarf  (WD)   stars  with  iron  -  rich,   core
composition. In this  connection, the present  paper is aimed  at
exploring the structure and evolution of iron - core WDs by means
of a detailed  and updated evolutionary  code. In particular,  we
examined  the  evolution  of  the  central  conditions,  neutrino
luminosity, surface gravity, crystallization, internal luminosity
profile and ages. We find that  the evolution of iron - rich  WDs
is  markedly  different  from  that  of  their  carbon  -  oxygen
counterparts. In particular, cooling is strongly accelerated  (up
to a  factor of  five for  models with  pure iron composition) as
compared with  the standard  case. Thus,  if iron  WDs were  very
numerous, some of  them would have  had time enough  to evolve at
lower luminosities than that corresponding  to the fall - off  in
the observed WD luminosity function.

\end{abstract}

\begin{keywords} stars: evolution - stars: interiors - stars:
white dwarfs

\end{keywords}

\section{Introduction}

Although the  theory of  electron degeneracy  is widely  accepted
amongst  the  astronomical  community,  up  to  very recently its
observational basis  was not  solid enough.  Obviously, the  most
useful objects  for testing  such a  theory are  white dwarf (WD)
stars. It is known that the structure of WDs is almost completely
dominated by electron degeneracy.  Theory of WD stars  predicts a
mass  -  radius  relation  (Chandrasekhar  1939)  that  should be
subject of observational test. Because of the great importance of
the  theory  of  electron  degeneracy  in  several  astrophysical
circumstances, great  effort has  been devoted  to improving  our
knowledge of the mass - radius relation for WD stars.

In this sense, recent observations carried out by the astrometric
satellite {\it Hipparcos} have allowed Provencal et al. (1998) to
substantially improve  the mass  and radius  determination for 20
WDs,  either  single  or  members  of  binary  systems. From very
accurate parallaxes,  these authors  determined precise  mass and
radius values,  without invoking  mass -  radius relations,  thus
making these WDs excellent targets for testing stellar degeneracy
directly. On the  basis of these  observations, Provencal et  al.
suggest in  particular that  at least  three objects  of their WD
sample appear to have an interior chemical composition consistent
with iron. Indeed, GD 140, EG50 and Procyon B have stellar  radii
that, for their observed  masses, are significantly smaller  than
those corresponding to a carbon - oxygen (CO) interior.  Needless
to say, such  results, if correct,  are clearly at  odds with the
standard  theory  of  stellar  evolution,  which  predicts  a  CO
interior for intermediate mass WDs.

It is nevertheless  worth noticing that  the only proposal  for a
physical process able  to account for  the formation of  iron WDs
is, to our knowledge, by means of explosive ignition of  electron
- degenerate ONeMg cores (Isern,  Canal \& Labay 1991). In  their
calculations, Isern et al.  find that, depending critically  upon
the ignition density and the velocity of the burning front,  such
an explosive ignition may lead to the formation of neutron stars,
thermonuclear supernovae or iron WDs.

Interestingly  enough,  the  implications  of  Provencal  et  al.
results about the possible existence of a population of iron  WDs
coupled to the  lack of modern  theoretical studies about  iron -
core  WDs  in  the  literature,  make  it worthwhile to perform a
detailed  exploration  of  the  structure  and  evolution of such
objects. As a matter of fact, to our knowledge the only study  of
the evolution of iron WDs was performed long ago by Savedoff, Van
Horn \&  Vila (1969);  however it  was based  on very  simplified
assumptions such  as the  neglect of  convection, crystallization
and electrostatic corrections to the equation of state. At  first
glance, one may think that  the evolution of these objects  could
be markedly different from  their CO counterparts. To  place this
suspicion on  a more  quantitative basis,  we have  carried out a
comprehensive study of the properties of iron - core WDs with the
emphasis placed on their evolution.

The  present   paper  is   organized  as   follows:  In   Section
\ref{sec:code}  we  briefly  describe  our  evolutionary code. In
Section \ref{sec:results} we summarise our main results. Finally,
Section  \ref{sec:conclusion}  is   devoted  to  discussing   the
implications  of  our  results  and  to  making  some  concluding
remarks.

\section{The Evolutionary Code} \label{sec:code}

The calculations we  present below were  performed with the  same
evolutionary code that we employed in our previous works on  WDs,
and we refer the reader to Althaus \& Benvenuto (1997, 1998)  for
details about both the  physical ingredients we incorporated  and
the  procedure  we  followed  to  generate the initial models. In
particular, the equation of state for the low - density regime is
that of  Saumon, Chabrier  \& Van  Horn (1995)  for hydrogen  and
helium plasmas, while the treatment for the high - density regime
(solid and liquid  phases) includes ionic  contributions, coulomb
interactions,  partially   degenerate  electrons,   and  electron
exchange and Thomas - Fermi contributions at finite  temperature.
The harmonic phonon contribution is that of Chabrier (1993).

High - density conductive opacities and the various mechanisms of
neutrinos emission for different chemical composition  ($^{4}$He,
$^{12}$C,  $^{16}$O,  $^{20}$Ne,  $^{24}$Mg, $^{28}$Si, $^{32}$S,
$^{40}$Ca and  $^{56}$Fe) are  taken from  the works  of Itoh and
collaborators (see  Althaus \&  Benvenuto 1997  for details).  In
addition   to   this,   we   include   conductive  opacities  and
Bremsstrahlung  neutrinos  for  the  crystalline  lattice   phase
following Itoh et  al. (1984a) and  Itoh et al.  (1984b; see also
erratum), respectively. The latter becomes relevant for WD models
with iron  core since,  as it  will be  clear later, these models
begin to develop a crystalline core at high stellar luminosities.

In Fig. \ref{fig:kcond}  we show the  conductive opacity at  some
selected temperature values for  iron, and for 50\% carbon  -
50\%  oxygen  plasmas.  The   downward  steps  are  due   to  the
crystallization of the  plasmas. Indeed, thermal  conductivity in
the  crystalline  phase  becomes  a  factor  2-4 smaller near the
melting temperature (see Itoh et  al. 1984a and Itoh, Hayashi  \&
Kohyama 1993 for  details), which, as  we shall see,  will affect
the rate of cooling of iron WD models.

We regard to neutrino  emission rates, we have  considered photo,
pair, plasma and Bremsstrahlung neutrino contributions. The total
emission rate is shown in  Fig. \ref{fig:enu} for the same  cases
as considered in Fig.  \ref{fig:kcond}. For a given  temperature,
the dominant neutrino emission process at low densities is  photo
neutrinos. At higher  densities there is  a bump in  the emission
rate  due  to  plasma  neutrinos,  whereas  at  higher  densities
Bremsstrahlung neutrinos take  over. Clearly, at  high densities,
neutrino  energy  losses  for  iron  plasmas  become  much   more
pronounced than for CO plasmas.

With respect to the energy transport by convection, we adopt  the
mixing length prescription usually  employed in most WD  studies.
Finally,  we   consider  the   release  of   latent  heat  during
crystallization  in  the  same  way  as  in  Benvenuto \& Althaus
(1997).

\section{ Evolutionary Results} \label{sec:results}

In this section we shall describe the main results we have  found
on the  evolution of  iron WDs.  We have  considered models  with
masses of \masa_msun=  0.50, 0.60, 0.70,  0.80, 0.90 and  1.0. In
view of the lack of a detailed theory about the formation of iron
WDs, we have taken into  account for each stellar mass  different
chemical stratifications. Specifically, we have adopted pure iron
cores embracing 99 (hereafter  referred to as pure  iron models),
75, 50 and  25 per cent  of the total  stellar mass plus  (in the
last  three  cases)  a  CO  envelope.  We  have also examined the
evolution of models  with a homogeneous  composition of iron  and
CO, by adopting a mass fraction for iron of 0.25, 0.50 and  0.75.
In  the  interests  of  comparison,  we  have  also  computed the
evolution  of  standard  CO  WD  models.  Since  standard  WDs of
different stellar masses are expected to have different  internal
composition of CO, we have adopted the chemical profiles  (kindly
provided  to  us  by  I.  Dom\'{\i}nguez)  resulting  from recent
evolutionary  calculations  of  WD  progenitors  (Salaris  et al.
1997). These profiles are also  considered in the CO envelope  of
our iron models.  In all the  cases just described  we have taken
into account  the presence  of an  outer helium  envelope  with
mass \m_mstar= 0.01  and adopted a  metallicity $Z$ of  0.001. In
the case of pure iron models, the transition from iron to  helium
layers  is  assumed  to  be  almost  discontinuous.  We have also
analysed the  effect of  a hydrogen  envelope on  iron models  by
including a  pure hydrogen  envelope with  \m_mstar= $10^{-5}$ on
the  top  of  the  helium  envelope  (in  this case we considered
$Z=0$).  Models  with  and  without  a  hydrogen envelope will be
hereinafter referred  to as  DA and  non -  DA, respectively. The
main  results  of  the  present  work  are  summarised  in  Figs.
\ref{fig:LnuTeff} to \ref{fig:FLumi3}. For completeness, we have
extended our  calculations to  the case  of stellar  mass of  0.4
\msun,  which  most  probably  would  have their origin in binary
systems.

We begin by examining Fig. \ref{fig:LnuTeff} in which we show the
neutrino luminosity in terms  of photon luminosity for  pure iron
and  CO  models.  As  expected,  high  mass  models  fade away in
neutrinos at higher luminosities than those corresponding to  low
mass models. Such behaviour is  found for both types of  interior
compositions  considered  here.  However,  because iron plasma is
more efficient neutrino emitter than CO (see Fig. \ref{fig:enu}),
note that for a given  mass and photon luminosity, iron  WDs have
appreciably higher  neutrino luminosity.  Consequently, neutrinos
is  the  dominant  energy  release  channel  for  iron  WDs up to
luminosities markedly lower than those corresponding to  standard
CO WDs.

The relation between central  temperature and density for  iron -
rich and  CO models  is shown  in Fig.  \ref{fig:TRhocent}. As we
have considered models  only in the  WD regime, these  curves are
almost vertical straight lines,  i.e. the density of  the objects
remained almost  constant during  the computed  stages. As  it is
well known, the higher the mass the higher the central density of
a WD, but note that for a fixed mass value, iron - rich WDs  have
central densities appreciably higher than the corresponding to CO
WDs. This is true even for models containing an iron core of only
25 per cent of the total  mass. The higher densities are a result
of basically two effects: One is the higher mean molecular weigth
- per - electron for  iron plasma ($\mu_{e}= 2.151344$ for  iron,
whereas $\mu_{e}=$ 2.000000 and  1.999364 for carbon and  oxygen,
respectively).  The  other  effect  is  that, because of the high
atomic number ($Z= 26$), iron plasma is subject to much  stronger
interactions than  a CO  one. Thus,  for a  given particle number
density, as  corrective terms  of the  pressure are  negative, an
iron  plasma  exerts  less  pressure,  forcing  a larger internal
density  than  for  CO  WDs.  It  is worth mentioning that as the
objects cool down, their internal temperature goes to zero. Thus,
the global structure of the objects asymptotically tends to  that
predicted by Hamada \& Salpeter (1961), as  expected\footnote{For
the sake of clarity, we do not include the results  corresponding
to the Hamada \& Salpeter zero temperature objects.}

Another important  characteristic of  the models  are their radii
and  surface  gravities   shown  in  Figs.   \ref{fig:RTeff1}  to
\ref{fig:gteeh} as  a function  of effective  temperature. Hot WD
models have larger  radii, due mainly  to the inflation  of their
non  -  and  partially  -  degenerate  outer  layers. It is quite
noticeable  that  iron  WDs  have  a  much smaller (higher) radii
(gravitational  acceleration)  than  their  CO counterparts. Note
also that,  for the  same stellar  mass and  iron content, models
with pure iron  cores are clearly  less compact than  homogeneous
iron models. These results can be compared with the observational
data  of  Provencal  et  al.  (1998)  to  infer the core chemical
composition of their  WD sample. In  particular, we add  in Figs.
\ref{fig:GTeff1} and \ref{fig:GTeff2} the observational data  for
the enigmatic case  of EG 50,  for which Provencal  et al. (1998)
quoted  a  surface  gravity  of  $  \log  g=  8.10  \pm 0.05$, an
effective temperature of 21700 K  $\pm$ 300 K and a  stellar mass
(derived without  relying on  a mass  - radius  relation) of 0.50
M$_{\odot}$ $\pm$ 0.02 M$_{\odot}$.  Having the stellar mass,  we
are in a position to estimate the core composition for this star.
In this context, we find that these values are compatible with  a
pure iron model  of $M \approx$  0.52 M$_{\odot}$. They  are also
consistent with homogeneous models with an iron abundance by mass
greater than $\approx$ 0.75. In Fig. \ref{fig:gteeh} we show  the
effect of a thick hydrogen envelope on our pure iron models.  The
presence  of  such  an  envelope  gives  rise to somewhat smaller
gravities,  an  oppossite  trend  to  what  is  needed to fit the
current observations of EG 50. For completeness, we show in  Fig.
\ref{fig:TcLumi} the central temperature versus photon luminosity
relation for some  selected models. Note  the change of  slope at
high luminosities, reflecting the end of neutrino dominated era.

As stated above, we  have also considered crystallization  in our
models. We assumed that  crystallization sets in when  the plasma
coupling  constant  $\Gamma$  reaches  the  value $\Gamma_m=180$,
where

\begin{eqnarray}    \Gamma =
2.275  \times 10^5\  {\rho^{1/3} \over  {T}} ({
{\mean{Z}}\over{\mean{A}}})^{1/3}
\mean{Z^{5/3}},
\end{eqnarray}

\noindent where $\mean{A}$  and $\mean{Z}$ denote,  respectively,
the averages  over abundances  by number  of the  atomic mass and
charge of  the different  species of  ions (see  Segretain et al.
1994). Our  choice of  $\Gamma_{m}$ value  is in  accordance with
studies carried out by Ogata \& Ichimaru (1987) and Stringfellow,
De Witt \&  Slattery (1990), and  it has been  used in recent  WD
evolutionary calculations  such as  Segretain et  al. (1994)  and
Salaris et al. (1997).

The growth of the crystal phase  in our models is shown in  Figs.
\ref{fig:XtalL1}  and  \ref{fig:XtalL2}  as  a function of photon
luminosity.  Very  large   differences  are  found   between  the
crystallization of iron WDs compared  to the case of standard  CO
WDs.   If   we   assume    a   pure   iron   composition,    then
$Z_{Fe}^2/A_{Fe}^{1/3}= 176.69$, whereas considering pure  carbon
and  oxygen  $Z_{C}^2  /  A_{C}^{1/3}$=  15.72  and  $Z_{O}^2   /
A_{O}^{1/3}$=  25.39,  respectively.  Thus,  iron  plasma reaches
crystallization conditions much earlier during the evolution.  In
addition, the  interior of  iron WDs  is much  denser than in the
standard case, as noted earlier. As a result, crystallization  of
iron WDs sets in at such high luminosities that neutrino emission
is the main  agent of energy  release. This is  in clear contrast
with  the  case  of  CO  WDs,  which  undergo  crystallization at
luminosities low enough that neutrino emission has already fadded
away. It should be noted that  for 1.0 \msun and 0.40 \msun  pure
iron models, the onset of crystallization occurs at  luminosities
2000 and 50  times higher than  the corresponding to  CO WDs with
the same mass. In the case of models with a pure iron core plus a
CO envelope, Fig. \ref{fig:XtalL1} depicts an interesting feature
worthy of comment. Indeed,  the weaker electrostatic coupling  of
CO plasma  as compared  to iron  plasma leads  to a  halt in  the
growth  of  the  crystalline  phase  once  the iron core has been
completely crystallized.  It is  only at  much later evolutionary
stages  that  crystallization  process  will  set in again. It is
worth  mentioning  that  in  the  case  of iron WDs, the analytic
treatment  of  the  growth  of  the  crystal  phase  presented in
Benvenuto \& Althaus (1995) is no longer valid, simply because in
that work it was assumed an isothemal interior.

As well known, the crystallization of the interior of WDs has two
effects on  the evolution.  One is  the release  of latent  heat,
which acts  as an  energy source  that delays  the evolution. The
other  is  the  change  of  the  specific heat at constant volume
$C_{v}$. In fact, ions are no longer free but subject to  undergo
small oscillations around  its corresponding lattice  equilibrium
position. In such a  case, $C_{v}= 3k D(\theta_{D}/T)$  where $D$
is the Debye function and $\theta_{D}= 1.74 \times 10^3\;  (2Z/A)
\rho^{1/2}$ is the Debye  temperature. Eventually, at low  enough
temperatures, $C_{v}  \rightarrow (T/\theta_{D})^{3}$.  Thus, the
crystallized interior has a  lower ability to store  heat, giving
rise to an acceleration of the cooling process.

In order to get a deeper insight on the role of these effects  in
the   evolution    of   iron    WDs,   we    present   in   Figs.
\ref{fig:profile06}  and   \ref{fig:profile10}  the   profile  of
interior luminosity  relative to  its surface  value at  selected
stages of the evolution of iron and CO WDs with 0.6 \msun and 1.0
\msun as a function of the fractional mass. Let us first  discuss
the  results  corresponding  to   the  0.6  \msun  models   (Fig.
\ref{fig:profile06}). In either of the iron and CO cases,  curves
labelled 1 correspond to  a evolutionary stage at  which neutrino
luminosity dominates,  giving rise,  as well  known, to  negative
luminosity values.  From then  on, in  the case  of CO WDs (lower
panel  of   Fig.  \ref{fig:profile06}),   neutrinos  fade   away,
producing a linear, smooth profile (corresponding to evolutionary
stages  limited  by  curves  2  and  3). Afterwards, the interior
crystallizes, as is reflected by the change of slope in curves 3,
4 and 5. Such a change of  slope is due to the release of  latent
heat. Note also the outwards direction of the propagation of  the
crystallization front as cooling proceeds.

Let us now discuss the  results corresponding to the case  of 0.6
\msun  iron  WD  (upper  panel  of  Fig. \ref{fig:profile06}). As
crystallization  occurs  very  early,  some  of  the evolutionary
stages we  have selected  correspond to  high luminosities. Curve
labelled  as  1 reaches negative  $L$  values because of neutrino
emission and curve 2 corresponds to a stage soon after the  onset
of  crystallization.  From  then  on,  the  crystal  front  moves
outwards during  the early  stages of  evolution. Note  the large
differences  found  in  the  luminosity  profile  in this case as
compared to the standard one,  largely due to the fact  that much
of the released latent heat is lost by neutrino emission.

In the case of 1.0 \msun models the differences are  dramatically
enhanced  (see  Fig.  \ref{fig:profile10}).  The  CO model (lower
panel) shows profiles very similar  to the previous CO case  with
the exception of the last stage of evolution included, for which,
as a  result of  the decrease  in the  specific heat $C_{v}$, the
profile is no longer linear. In the case of the iron model, curve
1   corresponds   to   a   stage   previous   to   the   onset of
crystallization. Curves 2, 3, 4  and 5 represent stages at  which
the crystal front moves outwards. Over these evolutionary stages,
latent heat is lost away  by neutrino emission. Finally, curve  7
corresponds to a stage so advanced that most of the luminosity is
provided by compression of the outermost layers. Indeed, most  of
the iron core has a very  low luminosity because of its very  low
specific heat ($\theta_{D}/T \approx 40$).

Let us  remind the  reader that  $C_{v}$ is  proportional to  the
number  of  particles.  Thus,  per  gram  of material, $C_{v}$ is
inversely proportional to the  atomic weight of the  constitutive
isotopes. In consequence,  for a given  stellar mass value,  iron
WDs  have  a  lower  total  capacity  of  storing  heat than that
corresponding to the standard  case in about (assuming  a mixture
of 50\% carbon and 50\% oxygen) $56/(0.5~ 12+ 0.5~ 16)= 4$. Thus,
it is clear that iron WDs should cool faster than CO ones, as  it
is shown  in Figs.  \ref{fig:AgeL1} -  \ref{fig:AgeL3}. In  these
figures we show the time spent by objects to cool down to a given
luminosity (at the  luminosity stages shown  in the figures,  our
election for the zero  age point, \ll_lsun=0, is  immaterial). We
find that in reaching  a given luminosity value,  low luminosity,
pure iron models has to evolve about a fifth of the time a CO  WD
need!. The  abrupt change  in the  rate of  cooling of  pure iron
models  (as  reflected  by  the  change  of  slope  in  the age -
luminosity relationship)  at the  high luminosity  range of these
figures is worthy of comment. In fact, it occurs when the crystal
front has just reached the outer layers of the iron core. Because
of the discontinuity of iron conductivity opacity at the  melting
temperature  (see  Fig.  \ref{fig:kcond}),  these  layers becomes
suddendly  much  more  transparent.  The  opacity of these layers
plays a  significant role  by regulating  the heat  flow from the
interior to  the outer  space; thus  such a  discontinuity in the
opacity is expected to affect the rate of cooling. The  situation
is more  clearly illustrated  by Fig.  \ref{fig:opacity} in which
the behaviour of opacity (conductive plus radiative) is shown  in
terms  of  the  outer  mass  fraction  for a pure iron model with
\masa_msun= 0.6 at different evolutionary stages. Note that  very
deep  in  the  star,   conduction  is  very  efficient,   so  the
discontinuity in the opacity will  play a minor role. It  is only
when crystallization reaches  the very outer  layers of the  iron
core that the  cooling rate will  be affected. This  explains the
fact that the induced effect  on the cooling times be  negligible
for models having only half  of their stellar masses composed  of
iron. The effect of a  hydrogen envelope on cooling of  pure iron
models  is  depicted  in  Fig.  \ref{fig:AgeL3}. As expected, the
presence of a hydrogen envelope increases the evolutionary  times
at very low luminosities. In part,  this is due to the excess  of
thermal energy the star has to get rid of when degeneracy reaches
the base  of the  convection zone,  thus producing  a bump in the
cooling curves (see D'Antona \& Mazzitelli 1989 for a  discussion
in the context of  CO WD models). We  should mention that in  the
present  calculations  we  have  not  investigated  the effect of
separation of carbon and oxygen during crystallization on the age
of  our  CO  models  (see  Salaris  et  al.  1997  and references
therein);  nor  did  we  take  into  account the effect of iron -
carbon phase  separation analysed  by Xu  \& Van  Horn (1992). In
view  of  the  fact  that  iron  -  rich  WDs crystallize at high
luminosities,  we  judge  that  the  induced delay in the cooling
times  of  our  iron  -  rich  models  brought  about by chemical
redistribution at  solidification would  be of  minor importance,
although more  details calculations  would be  required to  place
this assertion on a more quantitative basis.

Assuming a  constant birthrate  of WDs,  we have  computed single
luminosity  functions  (LF)  as  $dt/d$\ll_lsun  for  our  set of
models. The  results are  displayed in  Figs. \ref{fig:FLumi1} to
\ref{fig:FLumi3}. For clarity we  arbitrarily fixed the value  of
$dt/d$\ll_lsun= -5, -6 and -7  at \ll_lsun=0 for the two  sets of
iron and CO composition respectively. As in the previous figures,
the  differences  between  iron  and  CO  WDs are large. For iron
objects  at  stages  for  which  the  crystalline  phase is still
growing, the slope  of the LF  is rather larger  than for the  CO
case, especially for high mass objects. A striking feature  shown
by these  figures are  the spikes  characterizing the  LFs of the
pure iron sequences,  which are directly  understood in terms  of
the  discussion   presented  in   the  foregoing   paragraph.  As
explained, when the crystal front reaches the outer layers of the
iron  core,  they  become  suddendly  much more transparent (as a
result of the discontinuity of conductive opacity at the  melting
temperature,  see  Fig.  \ref{fig:opacity}),  giving  rise  to an
abrupt change in the rate of cooling of models, which  translates
into a discontinuity in the derivative of the evolutionary times.
Thus, LF shows a step downwards and then again increases steadily
up to the lowest $L$ considered here. We want to mention that the
behaviour of our theoretical  luminosity functions at the  lowest
luminosity values computed here may be affected by  extrapolation
of available opacity data.

\section{Discussion and conclusions} \label{sec:conclusion}

Motivated by recent observational evidence that seems to indicate
the existence of  white dwarf (WD)  stars with iron  - rich cores
(Provencal, et al. 1998), we have studied the evolution of iron -
core WDs.

In this paper we have constructed detailed evolutionary sequences
of   WDs   stars   with   different   chemical   stratifications.
Specifically,  we  have  computed  the  evolution  of models with
masses of \masa_msun= 0.40, 0.50, 0.60, 0.70, 0.80, 0.90 and  1.0
with pure iron cores embracing 99, 75, 50 and 25 per cent of  the
total stellar mass plus (in the last three cases) a CO  envelope.
We have also examined the evolution of models with a  homogeneous
composition of iron and CO, by adopting a mass fraction for  iron
of 0.25,  0.50 and  0.75. For  comparison purposes  with standard
results  we  have  also  computed  the  evolution of CO WD models
having the same masses. All of the models were assumed to have an
outer helium  envelope of  \m_mstar= 0.01,  and in  some cases we
analysed the effects of the  presence of a hydrogen envelope.  In
computing the structure  of such objects  we employed a  detailed
evolutionary  code  updated  to  account  for the physics of iron
plasmas properly.

In a set  of figures we  examined neutrino luminosities,  central
densities   and    temperatures,   radii,    surface   gravities,
crystallization,  internal  luminosity  profiles,  ages  and  the
luminosity function (at constant birthrate). Our results indicate
that iron  WDs evolve  in a  very different  way, as  compared to
standard CO WDs. These differences  are due to the fact  that the
mean molecular weight per electron for iron is higher than for CO
plasmas  and  also  to  the  stronger  corrections  to  the ideal
degenerate equation  of state  that causes  the pressure  of iron
plasmas to be below the values corresponding to the case of CO.

As  consequence  of  the  denser  interior, iron WDs have smaller
radii, greater surface gravities, higher internal densities, etc.
compared to standard  CO WDs of  the same mass.  We have compared
the predictions of our models with the current observational data
of the WD EG 50, for which Provencal et al. (1998) have suggested
an iron  - rich  composition. In  particular, we  found that this
object  is  consistent  with   WD  models  having  a   pure  iron
composition.  Likewise,  very  noticeable  are  the   differences
encountered in  the crystallization  process that  ocurrs at very
high  luminosities.  For  example,  the  onset of crystallization
occurs,  for  the  case  of  a 1  \msun  pure  iron   model, at a
luminosity 2000 times higher  than the corresponding to  the case
of a CO object with the same stellar mass.

Because iron particles  are much heavier  than carbon or  oxygen,
the specific  heat per  gram is  much lower,  indicating that the
interior  of  iron  WDs  is  able  to  store comparatively little
amounts of  heat. Thus,  it is  not surprising  that the  cooling
process at  very low  luminosities proceeds in a  much faster way
compared to the standard case.

We have  also computed  the single  luminosity function  for each
computed sequence. It is nevertheless worth noticing that, due to
the  uncertainties  present  in  the  birth  process, we have not
constructed an  integrated luminosity  function for  the computed
iron WD sequences. In any case it should be noticed that if  pure
iron  WDs,  to  which  class  EG  50  seems  to belong, were very
numerous, some of  them would have  had time enough  to evolve to
luminosities much  lower than  the corresponding  to the observed
fall - off of the WD luminosity function (\ll_lsun$\approx$ -4.5,
see Leggett,  Ruiz \&  Bergeron 1998  for details).  Thus, from a
statistical point  of view,  the lack  of a  tail in the observed
luminosity function strongly indicates  a low spatial density  of
pure iron WDs  and may be  employed to quantitatively  constraint
it.

Detailed tabulations of the  results presented in this  paper are
available upon request to the authors at their email addresses.

\section*{Acknowledgments}

We  are  deeply  acknowledged  to  our  anonymous  referee, whose
suggestions and comments greatly improved the original version of
this work. We are also grateful to I. Dom\'{\i}nguez for  sending
us the chemical profiles of her pre - white dwarf models.  O.G.B.
wishes to acknowledge to  Jan - Erik Solheim  and the LOC of  the
11th European Workshop on White Dwarfs held at Troms\/o  (Norway)
for  their  generous  support  that  allowed  him  to attend that
meeting were he  became aware of  the observational results  that
motivated the present work.

\newpage

%----------------------------------------------------------------
%---               FIGURE               CAPTIONS              ---
%----------------------------------------------------------------

\begin{figure}  \caption{Conductive  opacities  for  iron  (solid
line), and  for 50\%  carbon -  50\% oxygen  (short dashed lines)
plasmas. Curves correspond, from  bottom to top, to  temperatures
of  $5  \times  10^{6}$,  $10^{7}$,  $2 \times 10^{7}$, $5 \times
10^{7}$ and $10^{8}$ K respectively. Note the steps in the curves
that   correspond    to   crystallization    of   the    plasma.}
\label{fig:kcond} \end{figure}

\begin{figure}  \caption{Neutrino  energy  loss  rates  for  iron
(solid line), and for 50\% carbon  - 50\% oxygen (short dashed
lines)
plasmas. Curves correspond from bottom to top to temperatures  of
$10^{7}$, $2  \times 10^{7}$,  $5 \times  10^{7}$ and  $10^{8}$ K
respectively.  For  a  given  temperature,  at low densities, the
dominant neutrino emission process is due to photo neutrinos.  At
higher  densities  a  bump  in  the  emission rate appears due to
plasma neutrinos,  whereas at  highest densities,  Bremsstrahlung
neutrinos dominate.} \label{fig:enu} \end{figure}

\begin{figure}   \caption{Neutrino   luminosity   versus   photon
luminosity corresponding  to pure  iron -  (full lines)  and CO -
(dot  -  dashed  lines)  non  -  DA WD models with stellar masses
\masa_msun= 0.40, 0.50, 0.60, 0.70,  0.80, 0.90 and 1.0. For  the
sake of reference, the  line $L_{\gamma}=L_{\nu}$ is also  shown.
The higher  the model  mass, the  faster the  fadding in neutrino
luminosity. Clearly,  iron WDs  are much  more efficient neutrino
energy radiators. In fact, neutrino emission represents the  main
cooling agent  for iron  WDs up  to luminosities  significatively
lower than those corresponding to  the standard case of CO  WDs.}
\label{fig:LnuTeff} \end{figure}

\begin{figure}   \caption{Central   temperature   versus  central
density  for  iron  and  CO  non  -  DA  models  with  masses  of
\masa_msun= 0.40,  0.60, 0.80  and 1.0  (the higher  the mass the
higher the central density). For each stellar mass we depict  the
results corresponding to  CO models (dot  - dashed line),  and to
models with pure iron cores embracing  25, 50 and 75 per cent  of
the total stellar mass plus a CO envelope (dotted, short - dashed
and  long  -  dashed  lines,  respectively).  Finally, full lines
correspond to pure iron non - DA models. Notice that for a  fixed
mass  value,  iron  -  rich  WDs  have asymptotic central density
values  few  times  higher  than  the  corresponding  to CO WDs.}
\label{fig:TRhocent} \end{figure}

\begin{figure} \caption{Radii in  terms of effective  temperature
corresponding to  pure iron  (full lines)  and CO  (dot -  dashed
lines) non  - DA  models, and  to models  with a  pure iron  core
containing  half  of  the  total  stellar mass (dashed lines) for
masses \masa_msun=  0.40, 0.50,  0.60, 0.70,  0.80, 0.90  and 1.0
(the higher the mass the smaller the radius).} \label{fig:RTeff1}
\end{figure}

\begin{figure} \caption{Same as Fig. 5, but now dashed lines show
the results for homogeneous iron models with a iron abundance  by
mass of 0.5.} \label{fig:RTeff2} \end{figure}

\begin{figure}   \caption{Surface   gravities   versus  effective
temperature relation corresponding  to pure iron and CO  non - DA
models with masses of  \masa_msun= 0.40, 0.50, 0.60,  0.70, 0.80,
0.90  and  1.0  (the  higher  the  mass  the  higher  the surface
gravity). The meaning of the lines is the same as in Fig. 4.  The
observational data corresponding  to EG 50  are also included.  }
\label{fig:GTeff1} \end{figure}

\begin{figure} \caption{Same as Fig. 7 but for homogeneous  iron
models.} \label{fig:GTeff2} \end{figure}

\begin{figure}   \caption{Surface   gravity   versus    effective
temperature relation  corresponding to  pure iron  models of type
non  -  DA  and  DA  (solid  and dotted lines, respectively) with
masses of \masa_msun= 0.50, 0.60,  0.70, 0.80, 0.90 and 1.0  (the
higher the mass the higher the surface gravity). The inclusion of
a hydrogen envelope leads to somewhat smaller gravity values. The
observational data corresponding  to EG 50  are also included.  }
\label{fig:gteeh} \end{figure}

\begin{figure}  \caption{Central  temperature  as  a  function of
photon luminosity corresponding to pure iron (full lines) and  CO
(dotted lines) non - DA  models with masses of \masa_msun=  0.40,
0.50, 0.60, 0.70, 0.80, 0.90,  and 1.0. At low luminosities,  the
higher   the   mass   the   lower   the   central   temperature.}
\label{fig:TcLumi} \end{figure}

\begin{figure} \caption{Evolution of the crystallization front in
the   Lagrangian   coordinate   as   a   function  of  luminosity
corresponding to pure iron (full lines) and CO (dotted lines) non
- DA models, and to models with a pure iron core containing  half
of the total  stellar mass (dashed  lines) for stellar  masses of
(from right to  left) \masa_msun= 0.40,  0.50, 0.60, 0.70,  0.80,
0.90 and 1.0. Note  that iron models crystallize  at luminosities
far higher  that those  corresponding to  their CO  counterparts.
This is a result of the strong electrostatic coupling of the iron
plasma.  Note  also  that  once  the  iron  core  is   completely
crystallized,  crystallization  conditions  at  the CO plasma are
reached    only    at    much    later    evolutionary   stages.}
\label{fig:XtalL1} \end{figure}

\begin{figure} \caption{Same  as Fig.  11, but  now dashed  lines
show  the  results  for  homogeneous  iron  models  with  an iron
abundance by mass of 0.5.} \label{fig:XtalL2} \end{figure}

\begin{figure} \caption{Profile of the relative luminosity versus
the fractional mass for non - DA WDs models with 0.6 \msun having
pure iron (upper panel) and  CO (lower panel) interiors. For  the
case  of  the   iron  models  we   have  included  the   profiles
corresponding  to  luminosities  of  \ll_lsun=  -0.9004, -1.4154,
-1.6554,  -1.8588,  -2.0301  and  -4.9643  labelled  from  1 to 6
respectively;  whereas  for   CO  models  the   luminosities  are
\ll_lsun= -0.67,  -0.95, -3.49,  -3.65, -4.00  and -4.90 labelled
from   1   to   6   respectively.   For   details,   see   text.}
\label{fig:profile06} \end{figure}

\begin{figure} \caption{Profile of the relative luminosity versus
the fractional mass for non - DA models with 1.0 \msun WDs having
pure iron (upper panel) and  CO (lower panel) interiors. For  the
case  of  the   iron  models  we   have  included  the   profiles
corresponding  to  luminosities  of  \ll_lsun=  1.58447,  0.1668,
-0.1771, -0.5043, -0.9652, -3.8293 and -4.9036 labelled from 1 to
7  respectively;  whereas  for  CO  models  the  luminosities are
\ll_lsun= -0.12,  -0.65, -2.86,  -3.05, -3.44  and -5.00 labelled
from   1   to   6   respectively.   For   details,   see   text.}
\label{fig:profile10} \end{figure}

\clearpage

\begin{figure}    \caption{Age    versus    luminosity   relation
corresponding to pure iron (full lines) and CO (dotted lines) non
- DA models, and to models with a pure iron core containing  half
of the total  stellar mass (dashed  lines) for stellar  masses of
\masa_msun= 0.40, 0.50, 0.60, 0.70, 0.80, 0.90 and 1.0. Note that
pure iron models cool down about five times faster than  standard
CO WDs. At  the high luminosities  of this figure,  the change in
the slope of pure iron models is a result of the discontinuity of
conductive opacity  at the  crystallization front.  See text  for
details. } \label{fig:AgeL1} \end{figure}

\begin{figure} \caption{Same  as Fig.  15, but  now dashed  lines
show  the  results  for  homogeneous  iron  models  with  an iron
abundance by mass of 0.5.} \label{fig:AgeL2} \end{figure}

\begin{figure}    \caption{Age    versus    luminosity   relation
corresponding to pure iron (full lines) and CO (dotted lines)  DA
models for stellar masses of \masa_msun= 0.40, 0.50, 0.60,  0.70,
0.80, 0.90 and 1.0 (for iron models the minimum mass value  shown
is 0.5 $M_{\odot}$ ). Note  that pure iron models cool  down about
five  times  faster  than  standard  CO  DA  WDs.  At  the   high
luminosities of this figure, the change in the slope of pure iron
models is a result of the discontinuity of conductive opacity  at
the    crystallization    front.    See    text   for   details.}
\label{fig:AgeL3} \end{figure}

\begin{figure}  \caption{Opacity  in  terms  of  the  outer  mass
fraction for pure iron DA models with 0.6 \masa_msun. From top to
bottom,  curves  correspond  to  different  evolutionary   stages
characterized by \ll_lsun= -1.31, -1.78, -1.95, -2.23, -2.35  and
-2.71. Note  the change  of iron  opacity at  the crystallization
front. At $M_r/M_*$  =0.99, the change  in the opacity  is due to
the presence  of  helium  in  the  outer  layers} \label{fig:opacity}
\end{figure}

\begin{figure} \caption{Single  luminosity functions  (normalized
at \ll_lsun=  0) in  terms of  surface luminosity  for pure  iron
(full lines) and CO (dotted lines) non - DA models, and to models
with a pure iron core  containing half of the total  stellar mass
(dashed  lines)  for  stellar  masses  with  (from bottom to top)
\masa_msun= 0.40, 0.50, 0.60, 0.70,  0.80, 0.90 and 1.0. For  the
sake of clarity we have set arbitrarily LF= -5, -6 and -7 for the
three set of curves. For the  spikes in the iron LFs, see  text.}
\label{fig:FLumi1} \end{figure}

\begin{figure} \caption{Same  as Fig.  19, but  now dashed  lines
show  the  results  for  homogeneous  iron  models  with  a  iron
abundance by mass of 0.5.} \label{fig:FLumi2} \end{figure}

\begin{figure} \caption{Single  luminosity functions  (normalized
at \ll_lsun=  0) in  terms of  surface luminosity  for pure  iron
(full lines) and CO (dotted  lines) DA models for stellar  masses
with (from  bottom to  top) \masa_msun=  0.50, 0.60,  0.70, 0.80,
0.90 and 1.0. For the sake of clarity we have set arbitrarily LF=
-5 and -6 for the two set  of curves. For the spikes in the  iron
LFs, see text.} \label{fig:FLumi3} \end{figure}

\bsp

\label
{lastpage}


\begin{thebibliography}{99}

\bibitem{} Althaus L. G., Benvenuto O. G., 1997, ApJ, 477, 313

\bibitem{} Althaus L. G., Benvenuto O. G., 1998, MNRAS, 296, 206

\bibitem{} Benvenuto O. G., Althaus L. G., 1995, ApSS, 234, 11

\bibitem{} Benvenuto O. G., Althaus L. G., 1997, MNRAS, 288, 1004

\bibitem{} Chabrier G., 1993, ApJ, 414, 695

\bibitem{} Chandrasekhar S., 1939, An Introduction to the  Study
of Stellar Structure, Univ. of Chicago Press

\bibitem{} D'Antona F., Mazzitelli I., 1989, ApJ, 347, 934

%\bibitem{} D'Antona F., Mazzitelli I. 1990, ARA\&A, 28, 139

\bibitem{} Hamada T., Salpeter E. E., 1961, ApJ, 134, 683

\bibitem{} Isern J., Canal R., Labay J., 1991, ApJ, 372, L83

%\bibitem{} Itoh N., Kohyama, Y. 1983, ApJ, 275, 858

\bibitem{} Itoh N., Hayashi H, Kohyama Y. 1993, ApJ, 418, 405

\bibitem{} Itoh N., Kohyama Y., Matsumoto N., Seki M.,  1984a,
ApJ, 285, 758

\bibitem{} Itoh N., Kohyama Y., Matsumoto N., Seki M.,  1984b,
ApJ, 285, 304 and, 1987, ApJ, 322, 584 (erratum)

%\bibitem{}  Itoh  N.,  Adachi,  T.,  Nakagawa,  M., Kohyama, Y.,
%Munakata, H., 1989, ApJ, 339, 354

%\bibitem{} Itoh N.,  Mutoh, H., Hikita,  A., Kohyama, Y.,  1992,
%ApJ, 395, 622

%\bibitem{} Itoh N.,  Mutoh, H., Hikita,  A., Kohyama, Y.,  1993,
%ApJ, 404, 418

%\bibitem{} Koester  D., Chanmugam  G., 1990,  Rep. Prog. Phys.,
%53, 837

%\bibitem{} Munakata H., Kohyama Y., Itoh N., 1987, ApJ,  316,
%708 % partially degenerate electrons & Bremsstrahlung

\bibitem{} Leggett  S. K.,  Ruiz M.  T., Bergeron  P., 1998, ApJ,
497, 294

\bibitem{} Ogata S., Ichimaru S., 1987, Phys. Rev. A, 36, 5451.

\bibitem{} Provencal J. L.,  Shipman H. L., H\/og  E., Thejll
P., 1998, ApJ, 494, 759

\bibitem{} Salaris M.,  Dom\'{\i}nguez I., Garc\'{\i}a - Berro
E., Hernanz M., Isern J., Mochkovitch R., 1997, ApJ, 486, 413

\bibitem{} Saumon D.,  Chabrier G., Van  Horn H. M.,  1995, ApJS,
99, 713

\bibitem{} Savedoff M. P., Van  Horn H. M., Vila S. C., 1969,
ApJ, 155, 221

\bibitem{} Segretain L.,  Chabrier G., Hernanz M., Garc\'{\i}a -
Berro E., Isern J., Mochkovitch R., 1994, ApJ, 434, 641

\bibitem{} Stringfellow G. S.,  De Witt H. E., Slattery W. I.,
1990, Phys. Rev. A., 41, 1105

\bibitem{} Xu Z. M., Van Horn H. M., 1992, ApJ, 387, 662.

\end{thebibliography}
\end{document}